\documentclass[fp,twocolumn,letter]{jpsj3}
\usepackage{txfonts}
\usepackage{color}

\title{Metallic State in Rubidium-Loaded Low-Silica X zeolite}

\author{
Peter Jegli\v{c}$^{1}$\thanks{peter.jeglic@ijs.si},
Takehito Nakano$^2$,
Tadej Me\v{z}nar\v{s}i\v{c}$^{1}$,
Denis Ar\v{c}on$^{1,3}$,
and
Mutsuo Igarashi$^4$\thanks{igarashi@gunma-ct.ac.jp}
}

\inst{
$^1$Jo\v{z}ef Stefan Institute, Jamova 39, 1000 Ljubljana, Slovenia \\
$^2$Institute of Quantum Beam Science, Graduate School of Science and Engineering, Ibaraki University, Mito, Ibaraki 310-8512, Japan \\
$^3$Faculty of Mathematics and Physics, University of Ljubljana, Jadranska 19, 1000 Ljubljana, Slovenia \\
$^4$National Institute{black} of Technology, Gunma College, Toribamachi 580, Maebashi 371-8530, Gunma, Japan \\
}
   
\abst{\textcolor{black}{The ground state of alkali metals when the particle size decreases from bulk to nanometric atomic clusters is inevitably accompanied by quantum effects that can suppress their pristine metallic state.} 
We demonstrate that the metallic nature of rubidium clusters confined and arrayed in the framework of insulating low-silica X zeolite is preserved.
The $^{87}$Rb NMR spin-lattice relaxation assigned to rubidium clusters \textcolor{black}{in supercages} shows a Korringa behavior from 190~K down to 10~K, which is compatible with a macroscopic observation of low electrical resistivity.
The density of states at the Fermi level is found to be enhanced compared to the analogous sodium case, \textcolor{black}{consistent with a Holstein-Hubbard model of alkali-loaded zeolites.}
}

\begin{document}
\maketitle

Alkali metals at ambient conditions crystallize in body centered cubic structure\cite{Alkali_bulk_crystal}.
They are described as a textbook example of simple metals with a nearly-free electron model.
In the bulk, the interatomic electron interactions remove the degeneracy of alkali-metal $s$-states and form a conduction band, where the density of states (DOS) plays a crucial role in determining the properties of metals.
However, for small nanosized particles and atomic clusters of alkali atoms their properties are determined by the statistics of the $s$-electron-level distribution.\cite{Halperin_WP_1986,Klink_2000}  
Therefore, when alkali clusters are formed in \textcolor{black}{nanospaces or cages}, quantum effects become important and new electronic states emerge \textcolor{black}{with symmetries corresponding to} $1s$, $1p$ and $1d$ states, which are the solutions of the spherical quantum-well model.\cite{Nakano_2017}
These cluster states may be also influenced by electron correlations due to the Coulomb repulsion and electron-phonon interactions.
As a result, the metallic nature of alkali clusters becomes fragile and can be even fully suppressed. 

\par

Zeolites\cite{Breck_DW_1973} with a periodic array of empty voids called cages possess a framework, which is ideal to confine alkali clusters.
\textcolor{black}{In constrast to alkali nanoparticles formed on substrates, where the size distribution of these particles cannot be avoided, the zeolite frameworks offer regular arrangements of clusters with a well defined number of alkali atoms. \cite{Klink_2000}} 
In sodalite and zeolite A structures with relatively small cage sizes ($\sim$7~\AA~and $\sim$11~\AA~for $\beta$-cage and  $\alpha$-cage, respectively) and small sizes of windows between the adjacent cages ($\sim$3~\AA~and $\sim$5~\AA), the metallic nature of alkali clusters is strongly suppressed, leading to Mott insulating ground state, for which a variety of magnetic transitions were reported\cite{Nozue_Y_1992, Srdanov_VI_1998, Damjanovic_L_2000}.
On the contrary, larger cage and window sizes ($\sim$13~\AA~and $\sim$8~\AA~for a supercage) found in low-silica X (LSX) zeolite seem to preserve the metallic nature of alkali clusters when the cages are loaded with sodium\cite{Nozue_Y_2012}.
However, sodium loaded LSX zeolite is far from being a simple metal. 
Although the Drude term is observed in the optical spectra, resistivity remains relatively high, which is typical for bad metals\cite{Nakano_T_2010}. 
The $^{23}$Na spin-lattice relaxation rate divided by temperature, $1/T_1T$, shows a Korringa-type temperature independent behavior only at temperatures below 30 K, where a metallic ground state is concluded to exist on a microscopic scale with the small DOS at the Fermi level, $N(E_F)$. \cite{Igarashi_M_2016}

\par

To gain a better understanding of how the metallic state of alkali clusters is influenced by the confinement in LSX zeolite, it would be desirable to exchange sodium with heavier alkali atoms.
In the framework of the Holstein-Hubbard Hamiltonian ($t$-$U$-$S$-$n$ model),\cite{tUSn_1, tUSn_2} which captures all the essential ingredients of $s$-electrons of alkali clusters confined in zeolite cages, heavier alkali metals have lower ionization energies resulting in a shallower cluster potential.\cite{Nakano_2017}
This increases the $s$-electron transfer energy $t$, reduces the electron-phonon interaction energy $S$ and promotes the metallic state at otherwise comparable values of Coulomb repulsion energy $U$ and electron concentration $n$. 

\par

\textcolor{black}{
In this letter,} we report the first study of rubidium-loaded LSX zeolite employing complementary macroscopic and microscopic probes, and comparing it to the sodium-loaded LSX. 
Rb-form low-silica X zeolite with a chemical formula Rb$_{12}$Al$_{12}$Si$_{12}$O$_{48}$ (hereafter abbreviated as Rb$_{12}$-LSX), was obtained by ion exchange method from K-form LSX powder soaked in a RbCl aqueous solution.
Rb metal was adsorbed into the fully dehydrated Rb$_{12}$-LSX at a high loading density.
The average number of guest Rb atoms per supercage, $n\simeq 7.6$, was estimated from the weight ratio of the metal to Rb$_{12}$-LSX. 
The Rb-loaded sample is abbreviated as Rb$_n$/Rb$_{12}$-LSX. 
Similarly, Na-loaded Na-form LSX is abbreviated as Na$_n$/Na$_{12}$-LSX \cite{Nakano_T_2010}.

\par

The detailed experimental procedure for the dc resistivity, $\rho (T)$, measurement is explained elsewhere\cite{Kien}. 
Optical spectra were measured with a \textcolor{black}{Varian Cary 5G} optical spectrometer. 
Frequency-swept nuclear magnetic resonance (NMR) spectra were taken in a magnetic field of 9.39~T corresponding to the $^{87}$Rb Larmor frequency of $\nu^{87}_{\rm ref} = 130.895$~MHz, determined with a frequency standard aqueous solution of RbCl.
The spectra were measured by a solid echo pulse sequence with a pulse length of 2.4 $\mu$s.
The NMR spectra were taken at the frequency steps of 50 kHz and combined to obtain the full spectrum.
\textcolor{black}{
The temperature dependence of bulk magnetic susceptibility was measured in a magnetic field of 5~T with a Quantum Design MPMS-XL system.}

\begin{figure} [t!]
\centering
\includegraphics[width=0.90\linewidth]{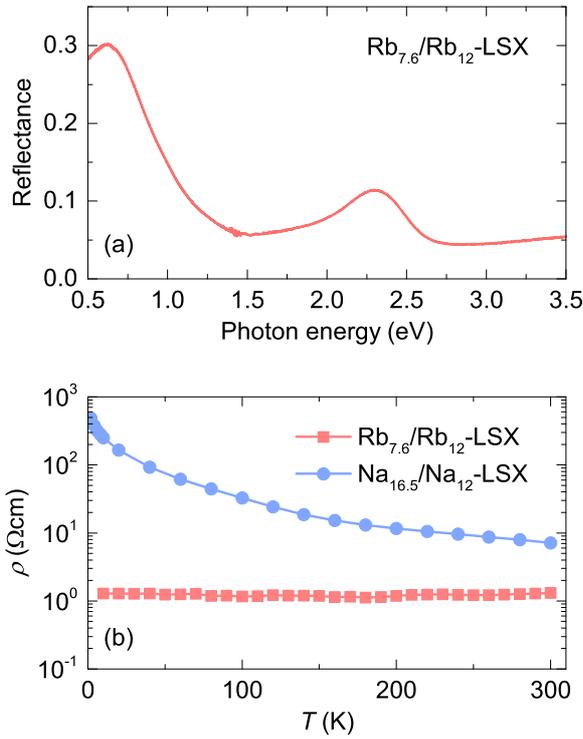} 
\caption{
\textcolor{black}{(a) Optical reflectance spectrum of Rb$_{7.6}$/Rb$_{12}$-LSX measured at room temperature.
(b) }Temperature dependences of dc electrical resistivity of Rb$_{7.6}$/Rb$_{12}$-LSX (red solid squares) and Na$_{16.5}$/Na$_{12}$-LSX zeolites (blue solid circles).
\textcolor{black}{The measured resistivity values are well above our measurement system threshold of $0.1~\Omega$cm due to the geometry constraints of the electrodes inside the sample cell.}}
\label{rho}
\end{figure}

\textcolor{black}{
Optical reflectance spectrum of Rb$_{7.6}$/Rb$_{12}$-LSX exhibits clear peaks at $\simeq$ 0.6~eV and $\simeq$ 2.3~eV (Fig.~\ref{rho}(a)). 
These peaks are attributed to the optical excitations of the $s$-electrons and confirm\cite{Kien} that
$s$-electrons provided by
guest Rb atoms are indeed confined in the supercage network and $\beta$-cages, respectively.}

\par

Fig.~\ref{rho}(b) shows that $\rho(T)$  of a pressed powder sample of Rb$_{7.6}$/Rb$_{12}$-LSX has more than \textcolor{black}{two orders of magnitude smaller value compared to the most conducting maximally loaded Na-form LSX at low temperatures.} 
It is important to note that $\rho(T)$ of Na$_{16.5}$/Na$_{12}$-LSX does not diverge and stays limited below $5\times10^2~\Omega$cm at 4~K, whereas Na$_{n}/$Na$_{12}$-LSX samples with moderate loading of guest sodium atoms showed divergent behavior with orders of magnitude higher resistivity, e.g. for $n=7.9$ the resistivity is above $10^7~\Omega$cm already at room temperature.\cite{Nozue_Y_2012}.
This was a clear indication of the insulator-to-metal crossover in sodium loaded  LSX zeolite at large $n$, which got further experimental support from NMR spin-lattice relaxation data.\cite{Igarashi_M_2016} 
Therefore, based on $\rho (T)$ it is reasonable to conclude that Rb$_{7.6}$/Rb$_{12}$-LSX \textcolor{black}{is positioned on the metallic side relative to the insulator-to-metal crossover compared to the Na$_{16.5}$/Na$_{12}$-LSX sample.}

\par

\textcolor{black}{
The observation of relatively high and almost temperature independent $\rho (T)$ in Rb$_{7.6}$/Rb$_{12}$-LSX is reminiscent of bad metal behavior, whereas Na$_{16.5}$/Na$_{12}$-LSX showing negative temperature coefficient of resistivity can be formally classified as bad insulator\cite{Dobrosavljevic}. 
In alkali-loaded zeolites, where the insulator-to-metal crossover can be successfully described by the $t$-$U$-$S$-$n$ model,\cite{Nakano_2017} a combination of strong electron correlations\cite{Aoki_1,Aoki_2} and electron-phonon interactions may explain the observed bad metal properties. 
The metal-to-insulator crossover is a result of a complex loading-level dependence of electric potential felt by the electrons confined to zeolite cages, where the electronic correlations and disorder both play an important role.\cite{Igarashi_M_2016}
In addition, the thermally activated motion of cations was observed in Na$_{n}/$Na$_{12}$-LSX,\cite{Igarashi_M_2013,Igarashi_JPSJ} which underlines the importance of electron-phonon interaction and polaron effects.\cite{Nakano_2017}  
}

\begin{figure} [t!]
\centering
\includegraphics[width=1.00\linewidth]{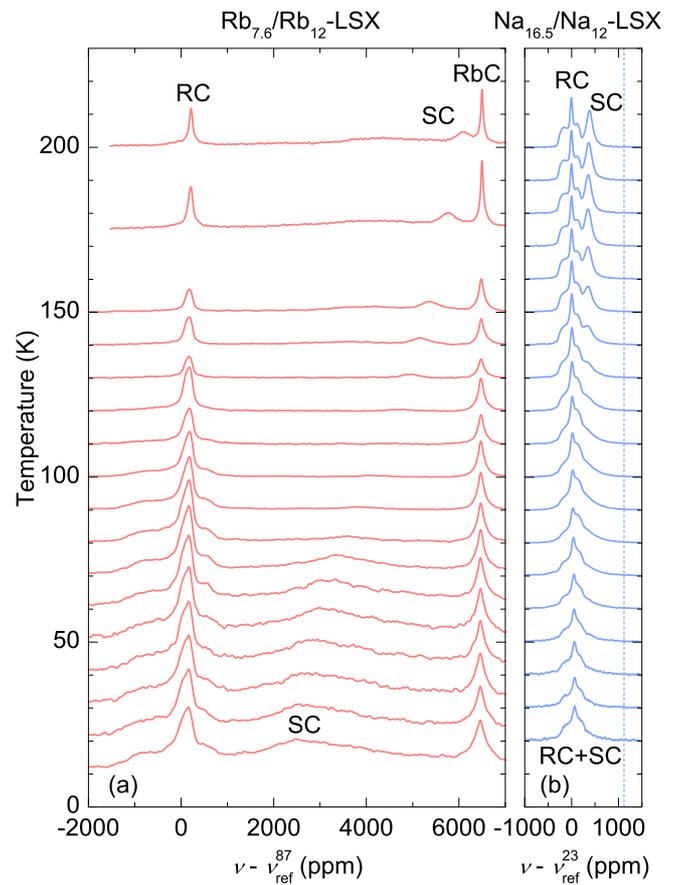}
\caption{\textcolor{black}{
(a) Temperature dependence of $^{87}$Rb NMR frequency-swept spectra of Rb$_{7.6}$/Rb$_{12}$-LSX in a magnetic field 9.39~T ($\nu^{87}_{\rm ref} = 130.895$~MHz). (b) $^{23}$Na NMR spectra of maximally loaded Na$_{16.5}$/Na$_{12}$-LSX measured in 4.7 T ($\nu^{23}_{\rm ref} = 52.9055$~MHz) are added for comparison\cite{Igarashi_M_2016} including new data below 100~K. The vertical dashed blue line marks the position of metallic sodium (NaC). RC, SC and  RbC denote the positions of residual, shifted and rubidium components, respectively.}}
\label{spectra}	
\end{figure}

\par

\begin{figure} [t!]
\centering
\includegraphics[width=0.90\linewidth]{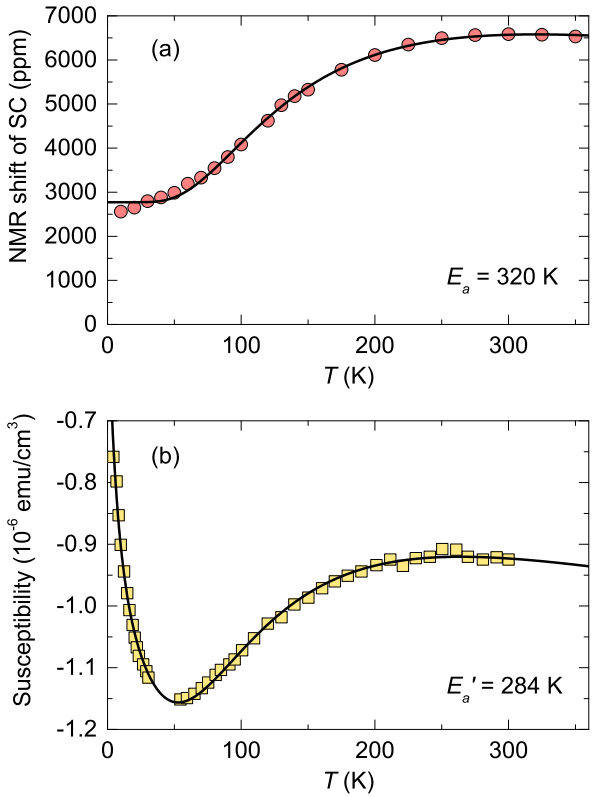}
\caption{\textcolor{black}{(a) Temperature dependence of $^{87}$Rb NMR shift at the position of SC in Rb$_{7.6}$/Rb$_{12}$-LSX. (b) Temperature dependence of bulk magnetic susceptibility of Rb$_{7.6}$/Rb$_{12}$-LSX. The negative temperature-independent diamagnetic contribution from a quartz tube has not been subtracted. The solid lines represent independent fits of experimental data using Eqs.~(\ref{Knight}) and (\ref{susc}), yielding comparable activation energies for small polarons of 320~K and 284~K, respectively. From the fits we obtain $K_{\rm iso}=2770$~ppm, $C_1=3.3\times10^6$~ppm$\cdot$K, $\chi_0=-1.3\times10^{-6}$~emu/cm$^3\cdot$K, $C_2=2.8\times10^{-4}$~emu/cm$^3\cdot$K, $C_3=7.7\times10^{-6}$~emu/cm$^3\cdot$K and $\theta=-9.2$~K. Please see text for more details.}}
\label{NMRshift}	
\end{figure}

\begin{figure} [t!]
\centering
\includegraphics[width=0.90\linewidth]{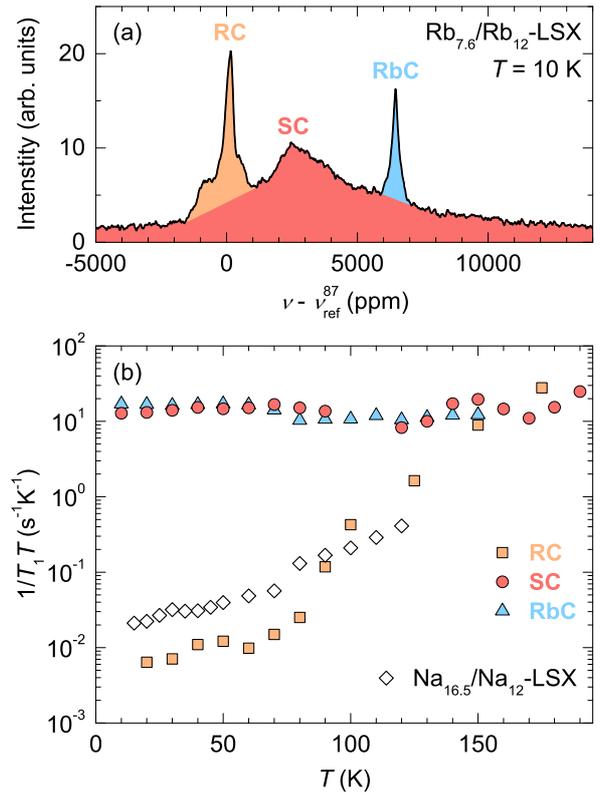}
\caption{(a) \textcolor{black}{Full $^{87}$Rb NMR frequency-swept spectrum at 10 K decomposed into residual (RC), shifted (SC) and rubidium (RbC) components. (b) $1/T_1T$ of Rb$_{7.6}$/Rb$_{12}$-LSX measured at the positions of RC, SC and RbC. $^{23}$Na $1/T_1T$ of Na$_{16.5}$/Na$_{12}$-LSX at the position where RC and SC are overlapping is added for comparison. The metallic sodium\cite{NaC} (not shown) has $1/T_1T = 0.21$~s$^{-1}$K$^{-1}$.}}
\label{NMR}	
\end{figure}

\textcolor{black}{
The $^{87}$Rb NMR spectra of Rb$_{7.6}$/Rb$_{12}$-LSX measured on cooling from 200~K to 10~K are shown in Fig.~\ref{spectra}(a).
Each spectrum is composed of three components, all of which have characteristic lineshapes and exhibit different temperature dependences of NMR shifts and spin-lattice relaxation time $T_1$.
The first component has a characteristic quadrupolarly-perturbed lineshape extending from -1600~ppm to 1600~ppm at low temperatures and its center is barely shifted with respect to $\nu^{87}_{\rm ref}$.
It can be assigned to the so-called residual component (RC) previously observed\cite{Igarashi_M_2013,Igarashi_M_2016} in the $^{23}$Na NMR study of Na-loaded Na-form LSX (Fig.~\ref{spectra}(b)).
The RC is attributed to rubidium sites located near the inner wall of the zeolite framework that are responsible for the framework charge compensation.}
The narrow component at 6400~ppm has an NMR shift corresponding to the metallic rubidium. \cite{Carter_GC_1977} 
It is natural to assume that this component originates from the residual metallic rubidium adsorbed at the surface of LSX crystals.
In this work we label it as the rubidium component (RbC).
\textcolor{black}{
The third component with a strong temperature dependence of NMR shift (Fig.~\ref{NMRshift}(a)) has all the features of the so-called shifted component (SC). At 10~K (Fig.~\ref{NMR}(a)) the SC is moderately shifted (2400~ppm), extremely broad (with full width at half maximum over 5000~ppm) and has by far the largest spectral weight.
Because of the cationic motion, its spin-spin relaxation rate (not shown) diverges at 110~K, which makes it difficult to observe the SC around this temperature. 
The SC has been previously observed in Na-loaded Na-form LSX and has become a hallmark of metallic behavior.\cite{Igarashi_M_2013,Igarashi_M_2016}
In Na$_{16.5}$/Na$_{12}$-LSX, the SC is clearly observed only above 130~K (Fig.~\ref{spectra}(b)), but it is overlapping with the RC at low temperatures.
This prevented the independent measurements of NMR shift and  $T_1$ at the SC originating from the Na clusters confined in supercages.
In contrast, we can follow the evolution of SC in the Rb$_{7.6}$/Rb$_{12}$-LSX sample down to low temperatures.}

\par

\textcolor{black}{
In analogy to the sodium case, the strong temperature dependence of $^{87}$Rb NMR shift at the position of SC (Fig.~\ref{NMRshift}(a)), can be well explained by a polaron model, where thermally activated behavior is associated with the creation/annihilation of localized small polarons from the bath of large (conducting) polarons.\cite{Igarashi_M_2013,Igarashi_M_2016}
At 10~K we can assume that the thermal excitations of large polarons are completely suppressed, so that the observed NMR shift of SC can be attributed to the isotropic Knight shift, $K_{\rm iso}$, which is proportional to the Pauli spin susceptibility of large polarons and is a measure of their DOS at the Fermi level, $N(E_F)$.
Thus the NMR shift at the position of SC can be modeled using the following empirical expression\cite{Igarashi_M_2013}
\begin{equation}
K(T)=K_{\rm iso}+\frac{C_1}{T} \exp \left(-\frac{E_a}{k_BT} \right) ,
\label{Knight}
\end{equation}
\noindent
where $E_a$, $k_B$ and $C_1$ are the activation energy for small polarons, Boltzmann constant and a shift prefactor, respectively.
The fit of experimental data using Eq.~(\ref{Knight}) gives $E_a=320$~K and $K_{\rm iso} \sim 2770$~ppm. The extracted activation energy is by a factor of $\sim 4$ smaller with respect to the value found in Na$_{16.5}$/Na$_{12}$-LSX.\cite{Igarashi_M_2013}}

\par

\textcolor{black}{
For completeness, we also report the bulk susceptibility data in Fig.~\ref{NMRshift}(b), which, as expected, display the same thermally activated behavior of small polarons with spin $S=1/2$. The susceptibility can be modeled with
\begin{equation}
\chi(T)=\chi_0+\frac{C_2}{T}\exp \left(-\frac{E'_a}{k_BT} \right) +\frac{C_3}{T-\theta},
\label{susc}
\end{equation}
\noindent
where $\chi_0$ is a sum of negative temperature-independent diamagnetism of quartz tube (not subtracted), spin susceptibility of large polarons and metallic rubidium contribution. $E'_a$ is the activation energy for small polarons obtained independently from the susceptibility data. $\theta$, $C_2$ and $C_3$ are the Weiss temperature and two independent constants, respectively.
The third term accounts for a Curie-Weiss contribution to susceptibility originating from a small amount of impurities.
Since NMR is a local technique, the latter are not detectable in NMR shift.\cite{Igarashi_M_2013}
Independent fit of bulk magnetic susceptibility using Eq.~(\ref{susc}) gives the activation energy $E'_a=284$~K, close to the value obtained from the NMR shift analysis, validating the polaron model.}

\par

\textcolor{black}{
The extracted isotropic Knight shift value of $K_{\rm iso} \sim 2770$~ppm should be compared to much larger value of 6400~ppm \cite{Carter_GC_1977}  found in metallic rubidium (Fig.~\ref{NMR}(a)).
We estimate that the $N(E_F)$ in Rb$_{7.6}$/Rb$_{12}$-LSX is by a factor of $\sim 2.3$ smaller than the corresponding value in bulk Rb.
Similarly we can analyse the situation in sodium case, where the SC is unfortunately overlapping with the RC and the corresponding $K_{\rm iso}$ can be only estimated to lie between 100~ppm and 250~ppm.
Since the metallic sodium has an isotropic Knight shift of 1120 ppm,  the $N(E_F)$ is reduced by a factor between 4.5 and 11 in Na$_{16.5}$/Na$_{12}$-LSX.
Based on the above crude estimates we can speculate that in Rb$_{7.6}$/Rb$_{12}$-LSX the rubidium clusters much more preserve the metallic ground state in comparison to the sodium case. 
}

\par

\textcolor{black}{
Now we move to the analysis of NMR spin-lattice relaxation data, which provides further support for the enhanced metallicity in rubidium case.
In Fig.~\ref{NMR}(b) we show $^{87}$Rb $1/T_1T$ measured at the RC, SC and RbC peak positions.
Magnetisation recovery curves were fitted with a stretched exponential model\cite{stretched} $\sim \exp [-(\tau/T_1)^\alpha]$ with a single effective $T_1$ to account for the distribution of $T_1$ due to the multiple rubidium sites contributing to RC and SC signals.\cite{Ikeda_T_2014}
The stretching exponents $\alpha$ for RC and SC are almost temperature independent between 10~K and 190~K with the corresponding values of 0.55 and 0.65, respectively.}
$1/T_1T$ of the RC shows a strong temperature dependence, which is due to the thermal motion of rubidium cations sitting in the LSX framework as already demonstrated in Na-form LSX.\cite{Igarashi_JPSJ}
\textcolor{black}{The same motion affects the RC lineshape too (Fig.~\ref{spectra}(a)), which becomes narrower above 100~K.}
As anticipated for a metallic rubidium, the RbC component shows a Korringa behavior in the whole temperature range.
Rather surprisingly, the SC shows an almost identical Korringa behavior from 190~K down to 10~K, which is a firm evidence of metallic ground state of rubidium clusters confined in the supercages of Rb$_{7.6}$/Rb$_{12}$-LSX.
\textcolor{black}{
Finally, the relative enhancement of the $N(E_F)$ in Rb$_{7.6}$/Rb$_{12}$-LSX compared to Na$_{16.5}$/Na$_{12}$-LSX, can be estimated from the ratio\cite{Igarashi_M_2016,NaC} 
\begin{equation}
\eta=\frac{\sqrt{^{87}T_1({\rm RbC})/^{87}T_1({\rm SC})}}{\sqrt{^{23}T_1({\rm NaC})/^{23}T_1({\rm SC})}},
\label{T1_DOS}
\end{equation}
\noindent
where $^{87}T_1(\rm RbC)$, $^{87}T_1(\rm SC)$, $^{23}T_1(\rm NaC)$ and $^{23}T_1(\rm SC)$ denote the $T_1$ values for metallic rubidium, rubidium SC, metallic sodium, and sodium SC, respectively.  
Plugging the corresponding low-temperature values into Eq.~(\ref{T1_DOS}) yields enhancement by a factor of  $\eta \sim 2.5$, which is in qualitative agreement with the above analysis of Knight shift and confirms the predicted dependence of $N(E_F)$ on alkali atom mass.
}


In this work, we have investigated the maximally Rb-loaded Rb-form LSX, which has been successfully synthesized for the first time. 
\textcolor{black}{Optical reflectance spectra indicate that loaded Rb atoms are confined in supercages and $\beta$-cages.}
\textcolor{black}{
At low temperatures the resistivity is by two orders of magnitude lower compared to the most conducting Na case, but still displays features typical for bad metals. 
The NMR shift  of SC and bulk magnetic susceptibility are fully compatible with the model of localised small polarons thermally activated from the bath of (conducting) large polarons.
The presence of polaron effects endorses the $t$-$U$-$S$-$n$ model used to describe the alkali-loaded zeolites and may, in combination with strong electron correlations, explain the observed bad metal behavior.
The $1/T_1 T$ of SC is temperature independent in the investigated temperature range, confirming the metallic ground state in Rb$_{7.6}$/Rb$_{12}$-LSX, despite the fact that rubidium clusters are formally confined in the insulating LSX framework.}
Whereas the maximally Na-loaded Na-form LSX zeolite shows that the DOS at the Fermi level is rather small compared to bulk sodium, the maximally Rb-loaded Rb-form LSX has a value comparable to bulk rubidium.
This corroborates with the $t$-$U$-$S$-$n$ model, which predicts that the metallic ground state becomes more stable for zeolites with heavier alkali clusters.\cite{Nakano_2017}

\begin{acknowledgments}
We thank Mr. T.~Goto and Prof. Y.~Nozue for their experimental support and fruitful discussion.
This study was partially supported by Grants-in-Aid for Scientific Research (KAKENHI) [Grant No. 16K05462(C)] from the Japan Society for the Promotion of Science.
\textcolor{black}{
This work was also supported by the Slovenian Research Agency (research core Grant No. P1-0125 and research projects No. J1-9145, No. N1-0052 and No. J2-8191).}
\end{acknowledgments}


\begin{thebibliography}{40}


\bibitem{Alkali_bulk_crystal}
C.~E.~Housecroft and A.~G.~Sharpe, {\it Inorganic Chemistry}
(Pearson, London 2018), 5th ed.

\bibitem{Halperin_WP_1986}
W.~P.~Halperin, 
Rev. Mod. Phys. \textbf{58}, 533 (1986). 

\bibitem{Klink_2000}
J.~J.~van der Klink and H.~B.~Brom, 
Prog. Nucl. Mag. Reson. Sp. \textbf{36}, 89 (2000). 

\bibitem{Nakano_2017}
T. Nakano and Y. Nozue, 
Adv. Phys. X \textbf{2}, 254 (2017). 

\bibitem{Breck_DW_1973}
D.~W.~Breck, {\it Zeolite molecular sieves: structure, chemistry, and use}
(Wiley, New York, 1973).

\bibitem{Nozue_Y_1992}
Y.~Nozue, T.~Kodaira, T.~Goto, 
Phys. Rev. Lett.  \textbf{68}, 3789 (1992).

\bibitem{Srdanov_VI_1998}
V.~I.~Srdanov, G.~D.~Stucky, E.~Lippmaa, G.~Engelhardt, 
Phys. Rev. Lett. \textbf{80}, 2449 (1998).

\bibitem{Damjanovic_L_2000}
L.~Damjanovi\'c, G.~D.~Stucky, V.~I.~Srdanov, 
J. Serb. Chem. Soc. \textbf{65}, 311 (2000).

\bibitem{Nozue_Y_2012}
Y.~Nozue, Y.~Amako, R.~Kawano, T.~Mizukane, T.~Nakano, 
J. Phys. Chem. Solids \textbf{73}, 1538 (2012).

\bibitem{Nakano_T_2010}
T.~Nakano, T.~Mizukane, Y.~Nozue, 
J. Phys. Chem. Solids \textbf{71}, 650 (2010).

\bibitem{Igarashi_M_2016}
M.~Igarashi, P.~Jegli\v{c}, A.~Krajnc, R.~\v{Z}itko, T.~Nakano, Y.~Nozue, D.~Ar\v{c}on,
Sci. Rep. \textbf{6}, 18682 (2016).
\textcolor{black}{
\bibitem{tUSn_1}
Y.~Shinozuka, 
J. Phys. Soc. Jpn. \textbf{56}, 4477 (1987).
}
\textcolor{black}{
\bibitem{tUSn_2}
P.~W.~Anderson, 
Phys. Rev. Lett. \textbf{34}, 953 (1975).
}
\bibitem{Kien} 
L.~M.~Kien, T.~Goto, D.~T.~Hanh, T.~Nakano, and Y.~Nozue, 
J. Phys. Soc. Jpn. \textbf{84}, 064718 (2015).
\textcolor{black}{
\bibitem{Dobrosavljevic}
V.~Dobrosavljevi\'{c}, N.~Trivedi, and J.~M.~Valles~Jr.,
{\it Conductor Insulator Quantum Phase Transitions}
(Oxford University Press, Oxford, 2012).}
\textcolor{black}{
\bibitem{Aoki_1}
H.~Aoki,
Appl. Surf. Sci. \textbf{237}, 2 (2004).}
\textcolor{black}{
\bibitem{Aoki_2}
R.~Arita, T.~Miyake, T.~Kotani, M.~van Schilfgaarde, T.~Oka, K.~Kuroki, Y.~Nozue and H.~Aoki,
Phys. Rev. B \textbf{69}, 195106 (2004).}
\bibitem{Igarashi_JPSJ} 
M.~Igarashi, P.~Jegli\v{c}, T.~Me\v{z}nar\v{s}i\v{c}, T.~Nakano, Y.~Nozue, N.~Watanabe, and D.~Ar\v{c}on,
J. Phys. Soc. Jpn. \textbf{86}, 075005 (2017).

\bibitem{Igarashi_M_2013}
M.~Igarashi, T.~Nakano, P.~T.~Thi, Y.~Nozue, A.~Goto, K.~Hashi, S.~Ohki, T.~Shimizu, A.~Krajnc, P.~Jegli\v{c}, D.~Ar\v{c}on, 
Phys. Rev. B \textbf{87}, 075138 (2013).

\bibitem{Carter_GC_1977}
G.~C.~Carter, L.~H.~Bennett, D.~J.~Kahan, 
{\it Metallic Shifts in NMR} 
(Pergamon Press, Oxford, 1977).

\bibitem{stretched}
D.~C.~Johnston, 
Phys. Rev. B \textbf{74}, 184430 (2006).
\textcolor{black}{
\bibitem{NaC}
R.~E.~Walstedt, {\it The NMR Probe of High-Tc Materials} (Springer Verlag, Berlin}

\bibitem{Ikeda_T_2014}
T.~Ikeda, T.~Nakano, Y.~Nozue, 
J. Phys. Chem. C \textbf{118}, 23202 (2014).

\end{thebibliography}
\end{document}